\documentclass[iop]{emulateapj}
\begin{document}
\submitted{Accepted for publication in The Astrophysical Journal}

\title{Detection of Strong Short-Term Variability in NGC 6946 X-1}

\author{Fengyun Rao\altaffilmark{1}, Hua Feng\altaffilmark{1}, \& Philip Kaaret\altaffilmark{2}}

\altaffiltext{1}{Department of Engineering Physics and Center for Astrophysics, Tsinghua University, Beijing 100084, China}
\altaffiltext{2}{Department of Physics and Astronomy, University of Iowa, Van Allen Hall, Iowa City, IA 52242, USA}

\shortauthors{Rao, Feng, \& Kaaret}
\shorttitle{Low Frequency Variability in NGC 6946 X-1}

\begin{abstract}

Using two archival {\it XMM-Newton} observations, we identify strong X-ray flux variations in NGC 6946 X-1 indicating it is the most variable ultraluminous X-ray source (ULX) on mHz time scales known so far. The 1-10~keV lightcurve exhibits variability with a fractional rms amplitude of 60\% integrated in the frequency range of 1-100~mHz. The power spectral density of the source shows a flat-topped spectrum that breaks at about 3~mHz with possible quasi-periodic oscillations (QPOs) near 8.5~mHz. Black hole binaries usually produce strong fast variability in the hard or intermediate state. The energy spectrum of NGC 6946 X-1 is dominated by two components, a 0.18~keV thermal disk and a power law with a photon index of $\approx$2.2, which is consistent with the intermediate state. The characteristic time scales of the X-ray emission suggests that the ULX may contain a black hole with a mass on the order of $10^{3}$ solar masses. 

\end{abstract}


\section{Introduction}
Ultraluminous X-ray sources (ULXs) are bright, nonnuclear X-ray sources found in nearby galaxies with luminosities above $3\times10^{39}$ erg s$^{-1}$ assuming isotropic emission, which is higher than the Eddington limit of a 20 $M_\sun$ black hole. Their nature remains a major astrophysical puzzle. They may be intermediate mass black holes of $10^2-10^4$~$M_\sun$ \citep{col99}, ordinary stellar mass black holes of $\sim$10~$M_\sun$ with beamed \citep{kin01,kor02} or super-Eddington radiation \citep{beg06}, or, massive stellar black holes of $\lesssim$100~$M_\sun$ formed in low metallicity environment with a high accretion rate \citep{zam09}. It is also possible that the high apparent luminosity is due to a combination of two or three of these factors \citep{pou07,kin08}. Due to the large distance and the brightness of the disk, dynamical measurement of the compact object mass via the motion of the companion is difficult and has not yet been accomplished for any ULX. X-ray spectroscopy could shed light on the black hole mass, but is only reliable in the thermal dominant state, which is rarely found in ULXs except one case firmly identified in M82 \citep{fen10b}. Another means to constrain compact object masses is X-ray timing. Characteristic time scales of X-ray emission, such as quasi-periodic oscillations (QPOs) and the break frequency in broad-band timing noise, have proved useful in constraining the mass of compact objects via a model independent calibration \citep[cf.][]{mch06}. 

Low-frequency QPOs have been detected in three ULXs, M82 X41.4+60 \citep{str03}, NGC 5408 X-1 \citep{str07}, and M82 X42.3+59 \citep{fen10}. To begin to understand these QPOs, an analogy with the QPOs detected from stellar mass black holes may be helpful. Low-frequency QPOs are classified into three main classes in stellar mass black holes, as type A, B, or C based on their amplitude, coherence, harmonics, red noise, etc.\ \citep[for a thorough discussion see][]{cas05}. The QPOs in M82 X41.4+60 and NGC 5408 X-1 are strong and narrow, and appear above strong flat-topped, band-limited noise (BLN) \citep{str03,fen07,str07}. The QPO frequencies vary in a wide range of 50-100~mHz for M82 X41.4+60 and 10-20~mHz for NGC 5408 X-1 and the BLN shows a break from a power-law index of 0 to $-1$ near the QPO frequency \citep{dew06,muc06,str09}. These properties are typical of the type C QPOs, which are the most common low-frequency QPOs found in Galactic black hole binaries usually appearing in the hard and intermediated states, and occasionally in the steep power-law state \citep{mcc06}. Recently, QPOs were discovered in another ULX in M82, X42.3+59. These QPOs show different characteristics: they are strong and broad, appear with very weak or absent red noise, detected only when the source is more luminous than $10^{40}$~erg~s$^{-1}$, and their central frequency varies in a narrow range from 3 to 4~mHz \citep{fen10}. These properties are unlike from type-C QPOs, but instead, are similar to those of type A/B QPOs, which show up in the steep power-law state. However, a survey of ULXs implies that the majority do not show strong fast variability \citep{fen05}. From a sample of 16 bright ULXs, \citet{hei09} concluded that in some ULXs the variability is significantly suppressed compared to bright black hole binaries and active galactic nuclei. Thus, detection of characteristic time scales in ULXs provides clues to their masses, but appears only rarely. Therefore, it is of significant interest to search for fast variability from ULXs. Here, we report the detection of strong X-ray variability in the mHz range in NGC 6946 X-1.

NGC 6946 is a nearly face-on spiral galaxy at a distance of 5.9~Mpc \citep{kar00}. The X-ray source NGC 6946 X-1 was first detected with Einstein \citep{fab87}, and later was found to be spatially associated with a nebula commonly referred to as MF 16 and thought to be a supernova remnant \citep{mat97} on the basis of ROSAT observations \citep{sch94,bla94}. Using Chandra observations, \citet{rob03} detected X-ray variability on time scales of minutes and argued that the X-rays arise from a black hole X-ray binary inside the nebula rather than extended emission from the nebula itself. The compact nature of the X-ray source was further strengthened with the detection of random variations in the X-ray flux on long time scales \citep{fri08}. The surrounding nebula MF 16 is thought to be a product of shock ionization \citep{bla01} plus unusually strong photoionization possibly originating from the central compact object \citep{abo08}. \citet{kaa10} detected an ultraluminous UV counterpart to the X-ray source right in the center of MF 16, which confirmed this scenario. Here, using two {\it XMM-Newton} observations available in the archive, we analyzed its timing and spectral behavior in \S~\ref{sec:ana}, and discussed the constraint on its physical nature in \S~\ref{sec:dis}.

\section{Observations and Data Analysis}
\label{sec:ana}

We examined all archival {\it XMM-Newton} observations of NGC 6946 with exposures longer than 20 ks to search for timing noise of X-1 at low frequencies. Among these observations, four out of six are not suited for this work because of high particle background, with $\sim$30\% background contamination in the source region. The remaining two observations (ObsIDs 0500730201 and 0500730101) were adopted for spectral and timing analysis.

In these two observations, both PN and MOS were operated in full frame mode. We used SAS 8.0.1 and calibration files current as of April 2010 for standard data reduction. For both the PN and MOS cameras, each energy spectrum was extracted from a $12\arcsec$-radius circular region to avoid the chip gap, and background from a nearby, circular, source-free region was subtracted. We fitted the spectra with the best model as discussed in \citet{rob03}: a cool multicolor disk plus a power-law component. All spectral parameters with errors are shown in Table~\ref{tab:ene}. The power-law component contributes $\sim$50\% to the total flux in the 0.3-10 keV band and $\sim$80\% in the 1-10 keV band. The two observations were made within a week and their spectral parameters are consistent within the 90\% errors. In particular, if we impose the same absorption column density on the two fits, almost identical results are obtained. Thus, we claim no significant spectral variation between the two observations, and, hence, we simultaneously fitted the two observations with a single model, as above, with the addition of a constant multiplying the flux normalization for ObsID 0500730101. As shown in the fourth column of Table~\ref{tab:ene}, this model fits well, with $\chi^2/{\rm dof} = 516.9/511$. The constant is estimated to be $0.98 \pm 0.03$, consistent with 1 within the 90\% error, which indicates an approximately constant flux.

\begin{deluxetable}{lccc}
\tablewidth{\columnwidth}
\tablecaption{Spectral Fits to the {\it XMM-Newton} Spectra \label{tab:ene}}
\tablehead{ObsID 						& 0500730201 			& 0500730101 		& Combined}
\startdata
Observation date 						& 2007-11-02 				& 2007-11-08			& \\
Exposure (ks) 							& 37	 					& 33 				& \\
$N_{\rm H}$ ($10^{22}$ cm$^{-2}$) 		& $0.35 \pm 0.05$			& $0.29 \pm 0.05$		& $0.32 \pm 0.03$\\
$T_{\rm in}$ (keV)						& $0.17 \pm 0.02$			& $0.20 \pm 0.03$		& $0.18 \pm 0.02$\\
$R_{\rm in}$ ($10^3$ km)				& $7.3^{+5.6}_{-2.3}$		& $4.2^{+3.3}_{-1.3}$	& $5.5^{+2.7}_{-1.4}$\\
$\Gamma$ 							& $2.33 \pm 0.13$			& $2.05 \pm 0.17$		& $2.19 \pm 0.10$\\
$f_{\rm X}$ ($10^{-13}$ erg/cm$^2$/s) 		& $8.3 \pm 0.3$			& $8.6 \pm 0.3$		& $8.5 \pm 0.2$\\
$L_{\rm X}$ ($10^{39}$ erg/s) 			& $11.7^{+1.3}_{-2.7}$		& $8.4^{+3.3}_{-1.2}$	& $10.0 \pm 1.6$\\
$L_{\rm d}$ ($10^{39}$ erg/s) 			& $5.9^{+3.2}_{-1.8}$		& $4.2^{+1.8}_{-0.9}$	& $5.0 \pm 1.4$\\
$\chi^2/{\rm dof}$ 						& $289.9/290$			& $221.5/217$		& $516.9/511$
\enddata
\tablecomments{The combined spectrum is described in the text. $N_{\rm H}$ is the absorption column density, $T_{\rm in}$ is the temperature of inner disk, $R_{\rm in}$ is the radius of the inner disk assuming a zero inclination angle, $\Gamma$ is the power-law photon index, $f_{\rm X}$ is the observed flux in 0.3-10~keV, $L_{\rm X}$ is the luminosity corrected for absorption in 0.3-10~keV, $L_{\rm d}$ is the disk luminosity corrected for absorption in 0.3-10~keV. Errors are quoted at a confidence of 90\%.}
\end{deluxetable}

\begin{figure}
\centering
\includegraphics[width=0.45\textwidth]{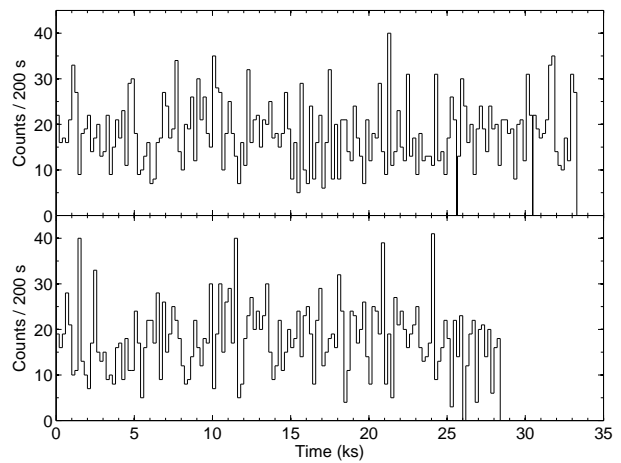}
\caption{X-ray lightcurve of NGC 6946 X-1 from PN data in the energy range of 1-10~keV for the observation 0500730201 ({\it top}) and 0500730101 ({\it bottom}). There are three GTIs in 0500730201 and two in 0500730101. The PSDs are created from the GTIs only.
\label{fig:lc}}
\end{figure}

\begin{figure}
\centering
\includegraphics[width=0.45\textwidth]{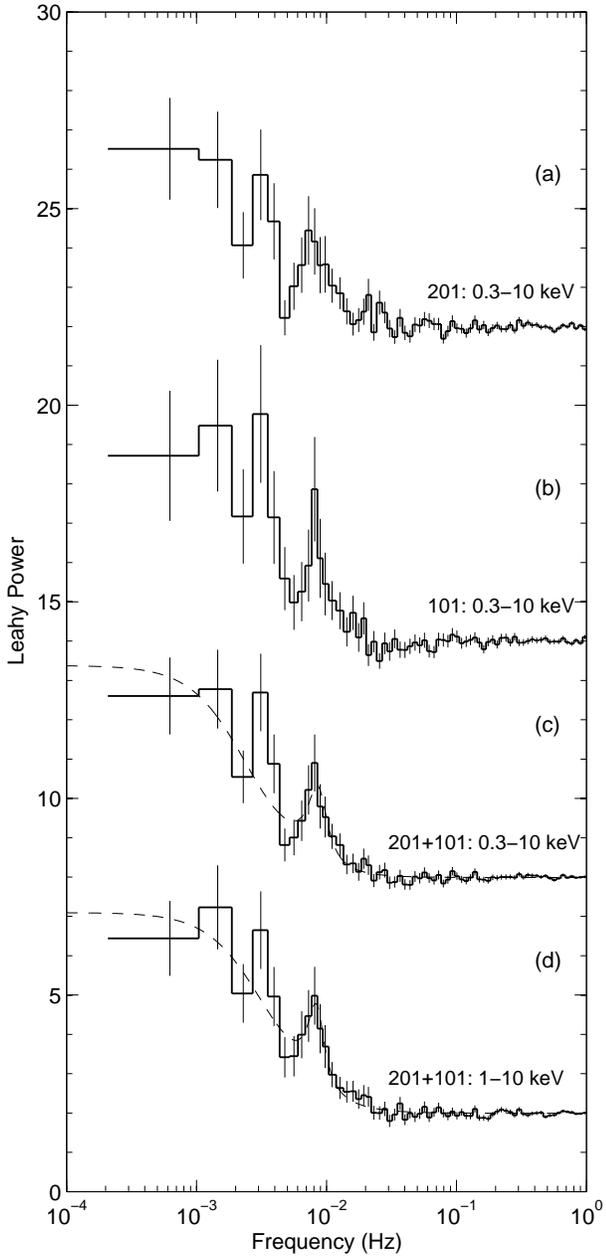}
\caption{X-ray power spectra of NGC 6946 X-1 calculated from the PN data. (a) PSD of observation 0500730201; (b) PSD of observation 0500730101; (c) PSD of the two observations; (d) similar to (c) but in 1-10 keV; The PSDs of (a)-(c) are vertically shifted for clarity by 20, 12, and 6, respectively. The dashed lines indicate the fits with the two Lorentzian plus constant model, with parameters listed in Table~\ref{tab:pow}.
\label{fig:pow}}
\end{figure}

Due to the large effective area and fine time resolution of PN, we created a light curve using PN events in the GTIs with a time step equal to the frame time ($\sim73.4$ ms). The PN lightcurves shown in Figure~\ref{fig:lc} have a time step of 200~s for visual inspection only. A Fourier transform was performed within each GTI with $2^{15}$ points and individual spectra were averaged to create the final one. We grouped the frequency channels linearly by a factor of 2 followed by a logarithmic binning by a factor of $10^{1/30}$. These two steps led to at least 26 and 20 power points for observations 0500730201 and 0500730101, respectively, combined into a single frequency bin as required to model the power spectrum using minimum $\chi^2$ fitting \citep{kli89,pap93,hei09}. The PSDs for observations 0500730201 and 0500730101 from photons in the energy range of 0.3-10~keV are shown in Figure~\ref{fig:pow}a and \ref{fig:pow}b, respectively, and are normalized following \citet{lea83}. They both show strong variability at low frequencies above the Poisson level, which is a constant value of 2 in the Leahy normalization. As for the energy spectra, we combined the PN power spectra from the two observations, see Figure~\ref{fig:pow}c. 

To quantify the variability we fitted the power spectra using XSPEC v12. First we fitted the 0.3-10~keV PSD with a power-law model for the red noise plus a constant component for the Poisson noise, which gives a $\chi^2$ of 142.5 with 88 degrees of freedom (dof). We then replace the power-law component with a broken power-law with a fixed index of zero below the break (a flat-topped model). This improves the fit to $\chi^2/{\rm dof} = 101.1/87$, corresponding to a chance probability of $5 \times 10^{-8}$ for the break as calculated using the F-test. A broad, zero-centered Lorentzian model is often used to characterize the flat-topped BLN \citep{bel02}. Fitting with a zero-centered Lorentzian plus a constant results in a fit much better than the unbroken power-law, $\chi^2/{\rm dof} = 98.8/88$, again confirming the presence of a break. Adding a narrow Lorentzian to account for excess noise around 6-10~mhz, possible QPOs, further improves the fit to $\chi^2/{\rm dof} = 77.7/85$, suggesting a positive detection with a chance probability of $1.3 \times 10^{-4}$. There are 91 independent frequency bins and the QPOs spreads over $\sim5$ bins, giving about 91/5 trials. Taking into account the number of trails, the chance probability is $2.3 \times 10^{-3}$ equivalent to a significance of 3$\sigma$. We note that this significance may be an underestimate, since the number of trails estimated in this way is best suited for periodic signals that concentrate all power in a single, independent frequency bin. Therefore, we conclude that we find a firm evidence for a break and a possible signature of QPOs in the power spectrum from NGC 6946 X-1. All timing parameters derived from the double Lorentzian model are listed in Table~\ref{tab:pow}.

\begin{deluxetable*}{lccccc}
\tablewidth{0pc}
\tablecaption{Timing parameters of NGC 6946 X-1 fitted with a double Lorentzian model \label{tab:pow}}
\tablehead{
\colhead{}    &  \multicolumn{2}{c}{PSD in 0.3-10~keV} &   \colhead{}   &
\multicolumn{2}{c}{PSD in 1-10~keV} \\
\cline{2-3} \cline{5-6} \\
\colhead{} & \colhead{Flat-topped Noise}   & \colhead{QPO}   &
\colhead{} & \colhead{Flat-topped Noise}   & \colhead{QPO}}
\startdata
$\nu$ (mHz)					& 0 					& $8.5 \pm 0.5$ 		&& 0 				& $8.3 \pm 0.4$ 		\\		
FWHM (mHz)					& $4.9 \pm 1.1$ 		& $3.8 \pm 1.5$ 		&& $6.8 \pm 1.6$		& $3.1 \pm 1.6$ 		\\
Norm ($10^{-2}$)	 			& $4.1 \pm 0.7$		& $1.1 \pm 0.3$		&& $5.3 \pm 0.9$		& $1.0 \pm 0.4$ 		\\
$\nu_{\rm max}$ (mHz) 			& $2.5 \pm 0.6$ 		& $8.7 \pm 0.7$ 		&& $3.4 \pm 0.8$ 		& $8.4 \pm 0.5$ 		\\
rms/mean ($\%$)				& $30.4 \pm 0.9$		& $26.4\pm 3.9$		&& $ 49.6\pm 1.4$ 	& $34.4 \pm 7.2$ 		\\
\noalign{\smallskip}\hline\noalign{\smallskip}
Possion level 					& \multicolumn{2}{c}{$1.988 \pm 0.003$} 		&& \multicolumn{2}{c}{$1.993 \pm 0.003$}		\\
$R_{\rm T}$ (counts s$^{-1}$)		& \multicolumn{2}{c}{0.1712} 					&& \multicolumn{2}{c}{0.0907}				\\
$R_{\rm B}$ (counts s$^{-1}$)		& \multicolumn{2}{c}{0.0033}					&& \multicolumn{2}{c}{0.0021}				
\enddata
\tablecomments{$\nu$, FWHM, and Norm are the centroid, width, and integrated area of a Lorentzian model in XSPEC, respectively. The units of `Norm' are Leahy power multiplied by Hz. $\nu_{\rm max}$ is the characteristic frequency where $\nu P_\nu$ peaks \citep{bel02}. The rms of flat-topped noise is calculated in 1-100~mHz frequency range. $R_{\rm T}$ and $R_{\rm B}$ are the total and background count rate in the source extraction region, respectively. Errors are quoted at a confidence level of 68.3\% (1$\sigma$).}
\end{deluxetable*}

PSDs at three independent energy bands, 0.3-1~keV, 1-2~keV, and 2-10 keV, were also produced following the same procedure, and the integrated rms in the frequency range of 1-100~mHz is $11\% \pm 3\%$, $53\% \pm 3\%$, and $70\% \pm 6\%$, respectively. This is similar to the behavior of black hole binaries where the timing noise amplitude often increases with energy. Since the variability appears mostly above 1 keV, we produced a PSD in the 1-10~keV energy range, as shown in Figure~\ref{fig:pow}d. However, the strength of the possible QPOs does not appear to increase with energy. New, long observations are needed to test the significance of the QPO. 

\section{Discussion}
\label{sec:dis}

Two {\it XMM-Newton} observations of NGC 6946 X-1 reveal strong X-ray variability in the energy range of 1-10~keV at short time scales with a band-limited spectrum that breaks at about 3 mHz and has a fractional rms amplitude of 60\% integrated over the range 1-100~mHz. We find evidence for possible QPOs with a central frequency of $\sim8.5$~mHz. The detection of strong fast variability confirms the compact nature of the source, and, together with the high luminosity, suggests that the ULX contains an accreting black hole.

A timing survey of bright ULXs indicates that only a small number of ULXs exhibit strong variability at time scales shorter than $10^{4}$~s \citep{hei09}. NGC 5408 X-1 and M82 X41.4+60 are the typical two. The integrated rms amplitude in 1-100~mHz range in the 1-8~keV band is around 50\% for NGC 5408 X-1 \citep{str09} and slightly lower for M82 X41.4+60 \citep{fen07}. The integrated rms of NGC 6946 X-1 in the same energy and frequency range is $59\% \pm 2\%$. Thus, NGC 6946 X-1 becomes the most noisy ULX reported so far. 

The timing noise amplitude of black hole binaries varies dramatically in response to the emission state. When thermal emission from the accretion disk dominates the energy spectrum, timing noise is largely suppressed. In contrast, strong variability is always expected in the hard state, when the Comptonization component is dominant, and often appears, with a relatively lower amplitude, in the steep power-law state when the Comptonization component is important \citep{mcc06}. This is also seen in ULXs. When M82 X41.4+60 changed its spectrum from a featureless power-law to a curved shape consistent with emission from a thermal disk, the previously detected QPOs and power continuum disappeared as well \citep{fen10b}. M82 X41.4+60 is believed to be in the hard state when the timing noise is seen based on its hard power-law spectrum \citep{fen10b}, while NGC 5408 X-1 is suggested to be in the steep power-law state due to its soft spectrum and the spectral-temporal correlation \citep{str09}. 

The extremely high variability of NGC 6946 X-1 suggests that the corona/Comptonization emission is very important and should be the dominant component similar to the hard or steep power-law state. With a fractional rms of 60\%, it is more likely to be in the hard or intermediate state rather than the steep power-law state. The energy spectrum of NGC 6946 X-1 from the two {\it XMM-Newton} observations is consistent with this scenario: the power-law component is dominant with a contribution of about 80\% in 1-10~keV and the power-law photon index of $2.2 \pm 0.1$ is intermediate between those typical of the hard and steep power-law states. However, ULXs may not follow the general evolution pattern that most Galactic sources have. For example, GX 339--4 shows a correlation between the variation amplitude and the spectral hardness, and the variation amplitude exceeds 0.3 only when the source is hardest during the `right branch' on its hardness-intensity diagram \citep{bel05}. It is obviously difficult to interpret the energy spectrum of NGC 6946 X-1 as being extremely hard. Conversely, GRS 1915+105 has shown an irregular pattern between the hardness and the total rms amplitude. The source often exhibits strong variation up to a fractional rms of 70\% \citep{mor97,mun99}. In this sense, NGC 6946 X-1 is perhaps a better analog to GRS 1915+105.

The break frequency of the broad-band noise is thought to be inversely proportional to the mass of the compact objects \citep[e.g.][]{mar03}.Cygnus X-1 shows a similar break in the broad-band noise, from an index of 0 to $-1$, at a frequency that varies in the range of 0.04-0.4 Hz depending on emission states \citep{bel90}. The break frequency for NGC 6946 X-1 is about 3~mHz, which is 13-130 times lower than that in Cygnus X-1. This suggests that NGC 6946 X-1 harbors a 130-1300 $M_\sun$ black hole, assuming a compact object mass of $10M_\sun$ in Cygnus X-1 \citep{her95}. This estimate depends on the assumption that the low frequency break in the PSD scales linearly with black hole mass and extends well above $10M_\sun$.

The break frequency of broad-band noise is found to be tightly correlated with the QPO frequency in black hole binaries \citep{wij99}. The break and QPO frequencies reported here for NGC 6946 X-1 appear to lie on a low frequency extension of the correlation measured for stellar mass black holes, suggesting that similar mechanisms produce the low frequency variability on both Galactic black hole binaries and this ULX. On the basis of the high QPO amplitude and the presence of the flat-topped noise component, we classify the QPOs as of type C \citep{cas05}. An interesting correlation has been found between the frequency of type-C QPOs and the photon index of the power-law component in the energy spectrum, and it is suggested that the QPO frequency is inversely scaled with the black hole mass at a given photon index \citep{sha09}. For a photon index range of 2.1-2.3, the QPO frequency of GRS 1915+105 varies roughly between 1-2 Hz \citep{vig03}, which is 120-240 times higher than the frequency of the possible QPOs in NGC 6946 X-1. Thus, assuming a positive detection of type-C QPOs in NGC 6946 X-1 and a 14 $M_\sun$ black hole in GRS 1915+105 \citep{har04}, we derive a black hole mass of $(1-4) \times 10^3$~$M_\sun$ for NGC 6946 X-1. We caution that the QPO frequency-photon index relation has been verified only for stellar mass objects.

Stellar mass black holes and active galactic nuclei (AGNs) appear to lie on a `variability plane' defined by a relation between the black hole mass, accretion rate, and characteristic frequency in the PSD ($\nu_l$, frequency of the lower high-frequency Lorentzian component) \citep{mch06,kor07}. Due to the limited sensitivity, $\nu_l$ is not detected in ULXs. \citet{cas08} proposed a means to estimate the black hole mass in ULXs with type C QPOs via the variability plane by assuming a relation between the QPO frequency and $\nu_l$. Thus in the efficient accretion case, the black hole mass can be written as $\lg (M/M_\sun) \gtrsim 0.5 \lg(L_{\rm X}/0.1c^2) - 0.51 \lg \nu_{\rm QPO} - 7.9$, where the inequality is due to the unknown bolometric correction. Given an X-ray luminosity of $10^{40}$~erg~s$^{-1}$ and a QPO frequency of 8.5~mHz, the estimated mass is 1500~$M_\sun$ or slightly higher (due to bolometric correction). The `variability plane' is tested for a wide range of black hole masses. However, the correlation between $\nu_l$ and $\nu_{\rm QPO}$ has not been tested in ULXs and the unknown bolometric correction introduces additional uncertainty.

The energy spectrum of the source from the two {\it XMM-Newton} observations is similar to that reported by \citet{rob03} from Chandra data, best-fitted with a model consisting of a cool thermal disk component with $kT \approx 0.18$~keV plus a power-law tail with a photon index of $\sim2.2$. The best-fit inner radius of the accretion disk is several $10^3$~km, which gives an upper limit of $\sim$$10^3$~$M_\sun$ for a Schwarzschild black hole or $\sim$$10^4$~$M_\sun$ for an extreme Kerr black hole. This is consistent with the value derived from the characteristic time scales. However, this mass estimate depends on interpretation of the soft component disk emission and that there is debate about the correct spectral model to apply to ULXs.

To summarize, we have detected strong X-ray variability in NGC 6946 X-1, which becomes the most variable ULX known so far. A break in the broad band timing noise is firmly identified at a frequency near 3~mHz. A tentative signature of QPOs, likely of type C, is seen. The source is most likely in the intermediate state and has behavior analogous to that of NGC 5408 X-1 and some states of GRS 1915+105. Several different relations known to hold for stellar and/or super‎massive black holes, each with their own caveats and systematic uncertainties, suggest that the characteristic time scales of the PSD imply the presence of a black hole with a mass on the order of $10^3$~$M_\sun$ in NGC 6946 X-1.

\acknowledgments We thank the anonymous referee for helpful comments. HF acknowledges funding support from the National Natural Science Foundation of China under grant No.\ 10903004 and 10978001, the 973 Program of China under grant 2009CB824800, and the Foundation for the Author of National Excellent Doctoral Dissertation of China under grant 200935.

\end{document}